# Study on the applicability of Varshni potential to predict the mass-spectra of the quark-antiquark systems in a non-relativistic framework

Etido P. Inyang[1], Ephraim P. Inyang[1], Eddy S. William[1] and Etebong E. Ibekwe[2]

[1] Theoretical Physics Group, Department of Physics, University of Calabar, PMB 1115, Calabar, Nigeria

[2] Department of Physics, Akwa Ibom State University, Ikot Akpaden, P.M.B 1167, Uyo, Nigeria

*Corresponding author email: etidophysics@gmail.com*

*Abstract*

In this work, we obtain the Schrödinger equation solutions for the Varshni potential using the Nikiforov-Uvarov method. The energy eigenvalues are obtained in non-relativistic regime. The corresponding eigenfunction is obtained in terms of Laguerre polynomials. We applied the present results to calculate heavy-meson masses of charmonium $c\bar{c}$ and bottomonium $b\bar{b}$. The mass spectra for charmonium and bottomonium multiplets have predicted numerically. The results are in good agreement with experimental data and the work of other researchers.

## 1. Introduction

It is a well-known fact by many researchers that quarkonium systems provide deep insight into the essential description of quantum chromodynamics (QCD) theory, particle physics and standard model [1-3]. Quarkonia with heavy quark and antiquark and their interaction are well described by the Schrödinger equation (SE). The solution of this SE with potential is one of the most fundamental problems in quarkonium systems. It is noted that the potentials considered should take into account two important features of the strong interaction, namely, asymptotic freedom and quark confinement [4-8]. The most fundamental potential used in studying quarkonium system is the Cornell potential, also known as Killingbeck potential. Most researchers have carried out works with Cornell potential [9,10]. For instance, Vega and Flores [11] solved the Schrödinger equation with the Cornell potential using the variational method and supersymmetric quantum mechanics (SUSYQM). Ciftci and Kisoglu [12] addressed non-relativistic arbitrary $l$-states of quark-antiquark through the Asymptotic Iteration Method (AIM). The energy eigenvalues with any $l \neq 0$ states and mass of the massive quark-antiquark system (quarkonium) were gotten. An analytic solution of the N-dimensional radial Schrödinger equation with the mixture of vector and scalar potentials via the Laplace transformation method (LTM) was studied by [13]. Their results were employed to analyze the different properties of the heavy-light mesons. Al-Jamel and Widyan [14] studied heavy quarkonium mass spectra in a Coulomb field plus quadratic potential using the Nikiforov-Uvarov method. In their work, the spin- averaged mass spectra of heavy quarkonia in a Coulomb plus quadratic potential is analyzed within the non-relativistic Schrödinger equation. Al-Oun et al.[15] examine heavy quarkonia characteristics in the general framework of a non-relativistic potential model consisting of a Coulomb plus quadratic potential. Kumar and Chand [16],carried out an asymptotic study to the N-dimensional radial Schrödinger equation for the quark-antiquark interaction potential employing asymptotic iteration method (AIM). Ibekwe et al. [17] solved the radial SE with an exponential, generalized, anharmonic Cornell potential using the series expansion method. They applied the bound state eigenvalues to study the energy spectra for CO, NO, CH and $N_2$ diatomic molecules and the mass spectra of heavy quarkonium systems. Furthermore, Omugbe et al. [18] solved the SE with Killingbeck potential plus an inversely quadratic potential model. They obtained the energy eigenvalues and the mass spectra of the heavy and heavy-light meson systems. In addition, Ali et al.[19] studied the energy spectra of mesons and hadronic interactions using Numerov's method. Their solutions were used to describe the phenomenological interactions between the charm-anticharm quarks via the model. The model accurately predicts the mass spectra of charmed quarkonium as an example of mesonic systems. Inyang et al.[20] obtained the Klein-Gordon equation solutions for the Yukawa potential using the Nikiforov-Uvarov method. The energy eigenvalues were obtained both in relativistic and non-relativistic regime. They applied the results to calculate heavy-meson masses of charmonium and bottomonium.

The Varshni potential is greatly important with applications, cutting across nuclear physics, particle physics and molecular physics [21]. The Varshni potential takes the form;

$$V(r) = a - \frac{abe^{-\alpha r}}{r} \tag{1}$$

where $a$ and $b$ are potential strengths for Varshni potential, $\alpha$ is the screening parameter which controls the shape of the potential energy curve as shown in figure 1 and $r$ the inter nuclear separation. The Varshni potential is a short range repulsive potential energy that plays an important role in chemical, particle and molecular physics [22]. This potential is used generally to describe bound states of the interaction of systems and has been applied in both classical and molecular physics. The Varshni potential was studied by Lim using the 2-body Kaxiras-Pandey parameters. He observed that Kaxiras and Pandey used this potential to describe the 2 body energy portion of multi-body condensed matter [23].

Many researchers have studied heavy mesons with non-exponential type potential like the Cornell without considering the exponential type [11-16]. Therefore, we intend to investigate the SE with the exponential type potential (Varshni potential) in the framework of NU method to obtain the mass spectra of quark- antiquark systems. To the best of our knowledge this is the first time Varshni potential is being studied with the aim of determining the mass spectra of heavy quarkonia systems using the NU method.

The paper is organized as follows: In section 2, the Nikiforov-Uvarov (NU) method is reviewed. In section 3, the bound state energy eigenvalues and the corresponding eigenfunction are calculated. In section 4, the results are discussed. In section 5, the conclusion is presented.

## 2. A brief review of Nikiforov-Uvarov (NU) method

The NU method is used to transform Schrödinger -like equations into a second-order differential equation through a coordinate transformation $y = y(r)$, of the form [24-28]

$$\psi''(y) + \frac{\tilde{\tau}(y)}{\sigma(y)}\psi'(s) + \frac{\tilde{\sigma}(y)}{\sigma^2(y)}\psi(y) = 0 \tag{2}$$

where $\tilde{\sigma}(y)$ and $\sigma(y)$ are polynomials, at most second degree and $\tilde{\tau}(y)$

is a first-degree polynomial. From Eq. (2) we obtain exact solution by using the transformation.

$$\psi(y) = \phi(y)\chi(y) \tag{3}$$

This transformation reduces Eq. (2) into a hypergeometric-type equation of the form

$$\sigma(y)\chi''(y) + \tau(y)\chi'(y) + \lambda\chi(y) = 0 \tag{4}$$

The function $\phi(y)$ can be defined as the logarithm derivative

$$\frac{\phi'(y)}{\phi(y)} = \frac{\pi(y)}{\sigma(y)} \tag{5}$$

where $\pi(y)$ is at most a first-degree polynomial. The second part of the wave functions in Eq. (4) is a hypergeometric-type function obtained by Rodriguez relation:

$$\chi_n(y) = \frac{B_n}{\rho(y)} \frac{d^n}{dy^n}\left[\sigma^n(y)\rho(y)\right] \tag{6}$$

where $B_n$ is the normalization constant and $\rho(y)$ the weight function which satisfies the condition below;

$$\frac{d}{dy}(\sigma(y)\rho(y)) = \tau(y)\rho(y) \tag{7}$$

where also

$$\tau(y) = \tilde{\tau}(y) + 2\pi(y) \tag{8}$$

For bound solutions, it is required that

$$\frac{d\tau(y)}{dy} < 0 \tag{9}$$

With $\pi(y)$ and parameter $\lambda$, the eigenfunctions and eigenvalues can be obtained using the definition of the following function

$$\pi(y) = \frac{\sigma'(y) - \tilde{\tau}(y)}{2} \pm \sqrt{\left(\frac{\sigma'(y) - \tilde{\tau}(y)}{2}\right)^2 - \tilde{\sigma}(y) + k\sigma(y)} \tag{10}$$

and

$$\lambda = k + \pi'(y) \tag{11}$$

The value of $k$ can be obtained by setting the discriminant in the square root in Eq. (10) equal to zero. As such, the new eigenvalues equation can be given as

$$\lambda + n\tau'(y) + \frac{n(n-1)}{2}\sigma''(y) = 0, (n = 0, 1, 2, \ldots) \tag{12}$$

## 3. Approximate solutions of the Schrödinger equation with Varshni potential

The Schrödinger equation (SE) for two particles interacting via potential $V(r)$ in three dimensional space, is given by [29].

$$\frac{d^2 R(r)}{dr^2} + \left[\frac{2\mu}{\hbar^2}(E_{nl} - V(r)) - \frac{l(l+1)}{r^2}\right] R(r) = 0 \qquad (13)$$

where $l, \mu, r$ and $\hbar$ are the angular momentum quantum number, the reduced mass for the quarkonium particle, inter-particle distance and reduced plank constant respectively.

We carry out Taylor series expansion of the exponential term in Eq. (1) up to order three, in order to make the potential to interact in the quark-antiquark system and this yields,

$$\frac{e^{-\alpha r}}{r} = \frac{1}{r} - \alpha + \frac{\alpha^2 r}{2} - \frac{\alpha^3 r^2}{6} + \ldots \qquad (14)$$

We substitute Eq. (14) into Eq. (1) and obtain

$$V(r) = -\frac{B}{r} - Cr + Dr^2 + A \qquad (15)$$

where

$$\left.\begin{array}{l} A = a + ab\alpha, \quad B = ab \\ C = \dfrac{ab\alpha^2}{2}, \quad D = \dfrac{ab\alpha^3}{6} \end{array}\right\} \qquad (16)$$

Upon substituting Eq. (15) into Eq. (13), we obtain

$$\frac{d^2 R(r)}{dr^2} + \left[\frac{2\mu E_{nl}}{\hbar^2} + \frac{2\mu B}{\hbar^2 r} + \frac{2\mu Cr}{\hbar^2} - \frac{2\mu Dr^2}{\hbar^2} - \frac{2\mu A}{\hbar^2} - \frac{l(l+1)}{r^2}\right] R(r) = 0 \qquad (17)$$

In order to transform the coordinate from $r$ to $y$ in Eq. (17), we set

$$y = \frac{1}{r} \qquad (18)$$

This implies that the 2$^{nd}$ derivative in Eq. (18) becomes;

$$\frac{d^2 R(r)}{dr^2} = 2y^3 \frac{dR(y)}{dy} + y^4 \frac{d^2 R(y)}{dy^2} \qquad (19)$$

Substituting Eqs. (18) and (19) into Eq. (17) we obtain

$$\frac{d^2 R(y)}{dy^2} + \frac{2y}{y^2}\frac{dR(y)}{dy} + \frac{1}{y^4}\left[\frac{2\mu E_{nl}}{\hbar^2} + \frac{2\mu By}{\hbar^2} + \frac{2\mu C}{\hbar^2 y} - \frac{2\mu D}{\hbar^2 y^2} - \frac{2\mu A}{\hbar^2} - l(l+1)y^2\right] R(y) = 0, \qquad (20)$$

Next, we propose the following approximation scheme on the term $\frac{C}{y}$ and $\frac{D}{y^2}$.

Let us assume that there is a characteristic radius $r_0$ of the meson. Then the scheme is based on the expansion of $\frac{C}{y}$ and $\frac{D}{y^2}$ in a power series around $r_0$; i.e. around $\delta \equiv \frac{1}{r_0}$, up to the second order. This is similar to Pekeris approximation, which helps to deform the centrifugal term such that the modified potential can be solved by NU method [30].

Setting $x = y - \delta$ and around $x = 0$ it can be expanded into a series of powers as;

$$\frac{C}{y} = \frac{C}{x+\delta} = \frac{C}{\delta\left(1+\frac{x}{\delta}\right)} = \frac{C}{\delta}\left(1+\frac{x}{\delta}\right)^{-1} \tag{21}$$

which yields

$$\frac{C}{y} = C\left(\frac{3}{\delta} - \frac{3y}{\delta^2} + \frac{y^2}{\delta^3}\right) \tag{22}$$

Similarly,

$$\frac{D}{y^2} = D\left(\frac{6}{\delta^2} - \frac{8y}{\delta^3} + \frac{3y^2}{\delta^4}\right) \tag{23}$$

Substituting Eqs. (22) and (23) into Eq.(20), yields

$$\frac{d^2 R(y)}{dy^2} + \frac{2y}{y^2}\frac{dR(y)}{dy} + \frac{1}{y^4}\left[-\varepsilon + \alpha y - \beta y^2\right] R(y) = 0 \tag{24}$$

where

$$-\varepsilon = \left(\frac{2\mu E_{nl}}{\hbar^2} - \frac{2\mu A}{\hbar^2} + \frac{6\mu C}{\hbar^2 \delta} - \frac{12\mu D}{\hbar^2 \delta^2}\right), \quad \alpha = \left(\frac{2\mu B}{\hbar^2} - \frac{2\mu C}{\hbar^2 \delta^2} + \frac{16\mu D}{\hbar^2 \delta^3}\right)$$

$$\beta = \left(\gamma - \frac{2\mu C}{\hbar^2 \delta^3} + \frac{6\mu D}{\hbar^2 \delta^4}\right), \quad \gamma = l(l+1) \tag{25}$$

Comparing Eq. (24) and Eq.(2) we obtain

$$\tilde{\tau}(y) = 2y, \quad \sigma(y) = y^2, \quad \tilde{\sigma}(y) = -\varepsilon + \alpha y - \beta y^2 \\ \sigma'(y) = 2y, \quad \sigma''(y) = 2 \tag{26}$$

By substituting Eq. (26) into Eq. (10) it yields

$$\pi(y) = \pm\sqrt{\varepsilon - \alpha y + (\beta + k) y^2} \tag{27}$$

To determine $k$, we take the discriminant of the function under the square root, which yields

$$k = \frac{\alpha^2 - 4\beta\varepsilon}{4\varepsilon} \tag{28}$$

We substitute Eq. (28) into Eq. (27) and have

$$\pi(y) = \pm\left(\frac{\alpha y}{2\sqrt{\varepsilon}} - \frac{\varepsilon}{\sqrt{\varepsilon}}\right) \tag{29}$$

We take the negative part of Eq. (29) and differentiate, which yields

$$\pi'_-(y) = -\frac{\alpha}{2\sqrt{\varepsilon}} \tag{30}$$

By substituting Eqs. (26) and (29) into Eq.(8) we have

$$\tau(y) = 2y - \frac{\alpha y}{\sqrt{\varepsilon}} + \frac{2\varepsilon}{\sqrt{\varepsilon}} \tag{31}$$

Differentiating Eq. (31) we have

$$\tau'(y) = 2 - \frac{\alpha}{\sqrt{\varepsilon}} \tag{32}$$

By using Eq. (11), we obtain

$$\lambda = \frac{\alpha^2 - 4\beta\varepsilon}{4\varepsilon} - \frac{\alpha}{2\sqrt{\varepsilon}} \tag{33}$$

And using Eq. (12), we obtain

$$\lambda_n = \frac{n\alpha}{\sqrt{\varepsilon}} - n^2 - n \tag{34}$$

Equating Eqs. (33) and (34), the energy eigenvalues of Eq. (17) is given

$$E_{nl} = a(1+b\alpha) - \frac{3ab\alpha^3}{2\delta} + \frac{ab\alpha^3}{\delta^3} - \frac{\hbar^2}{8\mu}\left[\frac{\frac{2\mu ab}{\hbar^2} - \frac{3\mu ab\alpha^2}{\hbar^2\delta^2} + \frac{16\mu ab\alpha^3}{6\hbar^2\delta^2}}{n + \frac{1}{2} + \sqrt{\left(l + \frac{1}{2}\right)^2 - \frac{ab\mu\alpha^3}{\delta^3\hbar^2} + \frac{16\mu ab\alpha^3}{6\hbar^2\delta^4}}}\right]^2 \tag{35}$$

To determine the wavefunction, we substitute Eqs. (26) and (29) into Eq.(5) and obtain

$$\frac{d\phi}{\phi} = \left(\frac{\varepsilon}{y^2\sqrt{\varepsilon}} - \frac{\alpha}{2y\sqrt{\varepsilon}}\right)dy \tag{36}$$

Integrating Eq. (36), gives us

$$\phi(y) = y^{-\frac{\alpha}{2\sqrt{\varepsilon}}} e^{-\frac{\varepsilon}{y\sqrt{\varepsilon}}} \tag{37}$$

By substituting Eqs. (26) and (29) into Eq. (7) and integrating, thereafter simplify we obtain

$$\rho(y) = y^{-\frac{\alpha}{\sqrt{\varepsilon}}} e^{-\frac{2\varepsilon}{y\sqrt{\varepsilon}}} \qquad (38)$$

Substituting Eqs. (26) and (38) into Eq.(6) we have

$$\chi_n(y) = B_n e^{\frac{2\varepsilon}{y\sqrt{\varepsilon}}} y^{\frac{\alpha}{\sqrt{\varepsilon}}} \frac{d^n}{dy^n}\left[e^{-\frac{2\varepsilon}{y\sqrt{\varepsilon}}} y^{2n-\frac{\alpha}{\sqrt{\varepsilon}}}\right] \qquad (39)$$

The Rodrigues' formula of the associated Laguerre polynomials is

$$L_n^{\frac{\alpha}{\sqrt{\varepsilon}}}\left(\frac{2\varepsilon}{y\sqrt{\varepsilon}}\right) = \frac{1}{n!} e^{\frac{2\varepsilon}{y\sqrt{\varepsilon}}} y^{\frac{\alpha}{\sqrt{\varepsilon}}} \frac{d^n}{dy^n}\left(e^{-\frac{2\varepsilon}{y\sqrt{\varepsilon}}} y^{2n-\frac{\alpha}{\sqrt{\varepsilon}}}\right) \qquad (40)$$

where

$$\frac{1}{n!} = B_n \qquad (41)$$

Hence,

$$\chi_n(y) \equiv L_n^{\frac{\alpha}{\sqrt{\varepsilon}}}\left(\frac{2\varepsilon}{y\sqrt{\varepsilon}}\right) \qquad (42)$$

Substituting Eqs. (37) and (42) into Eq. (3) we obtain the wavefunction of Eq.(17) in terms of Laguerre polynomials as

$$\psi(y) = N_{nl} y^{-\frac{\alpha}{2\sqrt{\varepsilon}}} e^{-\frac{\varepsilon}{y\sqrt{\varepsilon}}} L_n^{\frac{\alpha}{\sqrt{\varepsilon}}}\left(\frac{2\varepsilon}{y\sqrt{\varepsilon}}\right) \qquad (43)$$

where $N_{nl}$ is normalization constant, which can be obtain from

$$\int_0^\infty |N_{nl}(r)|^2 dr = 1 \qquad (44)$$

**4. Results and discussion**

4.1 Results

The mass spectra of the heavy quarkonium system such as charmonium and bottomonium that have the quark and antiquark flavor is calculated and we apply the following relation [31,32]

$$M = 2m + E_{nl}, \qquad (45)$$

where $m$ is quarkonium bare mass, and $E_{nl}$ is energy eigenvalues. By substituting Eq. (35) into Eq. (45) we obtain the mass spectra for Varshni potential as:

$$M = 2m + a(1+b\alpha) - \frac{3ab\alpha^3}{2\delta} + \frac{ab\alpha^3}{\delta^3} - \frac{\hbar^2}{8\mu}\left[\frac{\frac{2\mu ab}{\hbar^2} - \frac{3\mu ab\alpha^2}{\hbar^2 \delta^2} + \frac{16\mu ab\alpha^3}{6\hbar^2 \delta^2}}{n + \frac{1}{2} + \sqrt{\left(l + \frac{1}{2}\right)^2 - \frac{ab\mu\alpha^3}{\delta^3 \hbar^2} + \frac{16\mu ab\alpha^3}{6\hbar^2 \delta^4}}}\right]^2 \qquad (46)$$

In order to test for the accuracy of the predicted results determined numerically, we used a Chi square function defined by [33]

$$\chi^2 = \frac{1}{n}\sum_{i=1}^{n}\frac{\left(M_i^{Exp.} - M_i^{Theo.}\right)}{\Delta_i} \qquad (47)$$

where $n$ runs over selected samples of heavy mesons, $M_i^{exp.}$ is the experimental mass of heavy-mesons, while $M_i^{Th}$ is the corresponding theoretical prediction. The $\Delta_i$ quantity is experimental uncertainty of the masses. Intuitively, $\Delta_i$ should be one. The tendency of overestimating Chi square value is that, it reflects some mean error per heavy meson state.

**Table1.** Mass spectra of charmonium in $(GeV)$ ($m_c = 1.488\ GeV$, $\mu = 0.744\ GeV$, $\alpha = -0.976$, $\delta = 1.700\ GeV$, $\hbar = 1$, $a = -48.049\ GeV$ and $b = 3.020\ GeV$)

| State | Present work | [30] | [20] | Experiment[35] |
|---|---|---|---|---|
| 1S | 3.096 | 3.096 | 3.096 | 3.096 |
| 2S | 3.686 | 3.686 | 3.672 | 3.686 |
| 1P | 3.295 | 3.255 | 3.521 | 3.525 |
| 2P | 3.802 | 3.779 | 3.951 | 3.773 |
| 3S | 4.040 | 4.040 | 4.085 | 4.040 |
| 4S | 4.269 | 4.269 | 4.433 | 4.263 |
| 1D | 3.583 | 3.504 | 3.800 | 3.770 |
| 2D | 3.976 | - | - | 4.159 |
| 1F | 3.862 | - | - | - |

**Table 2.** Mass spectra of bottomonium in $(GeV)$ ($m_b = 4.680\ GeV$, $\mu = 2.340\ GeV$, $\alpha = -0.952$, $\delta = 1.70\ GeV$, $\hbar = 1$, $a = -14.352\ GeV$ and $b = 3.084\ GeV$)

| State | Present work | [30] | [20] | Experiment[35] |
|---|---|---|---|---|
| 1S | 9.460 | 9.460 | 9.462 | 9.460 |
| 2S | 10.569 | 10.023 | 10.027 | 10.023 |
| 1P | 9.661 | 9.619 | 9.963 | 9.899 |
| 2P | 10.138 | 10.114 | 10.299 | 10.260 |
| 3S | 10.355 | 10.355 | 10.361 | 10.355 |
| 4S | 10.567 | 10.567 | 10.624 | 10.580 |
| 1D | 9.943 | 9.864 | 10.209 | 10.164 |
| 2D | 10.306 | - | - | - |
| 1F | 10.209 | - | - | - |

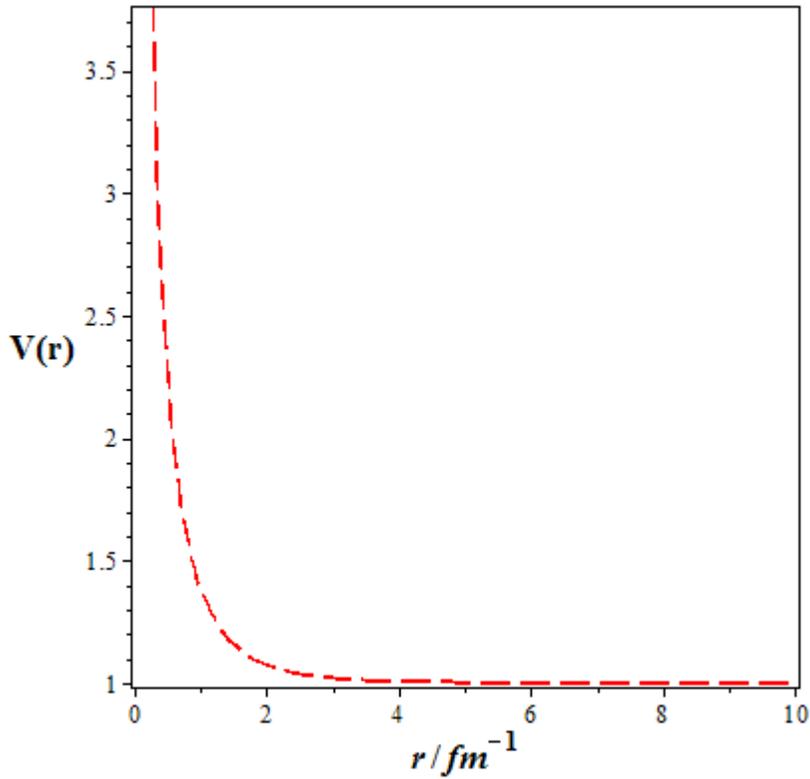

Fig.1. Plots of Varshni potential with r in ( $fm^{-1}$ )

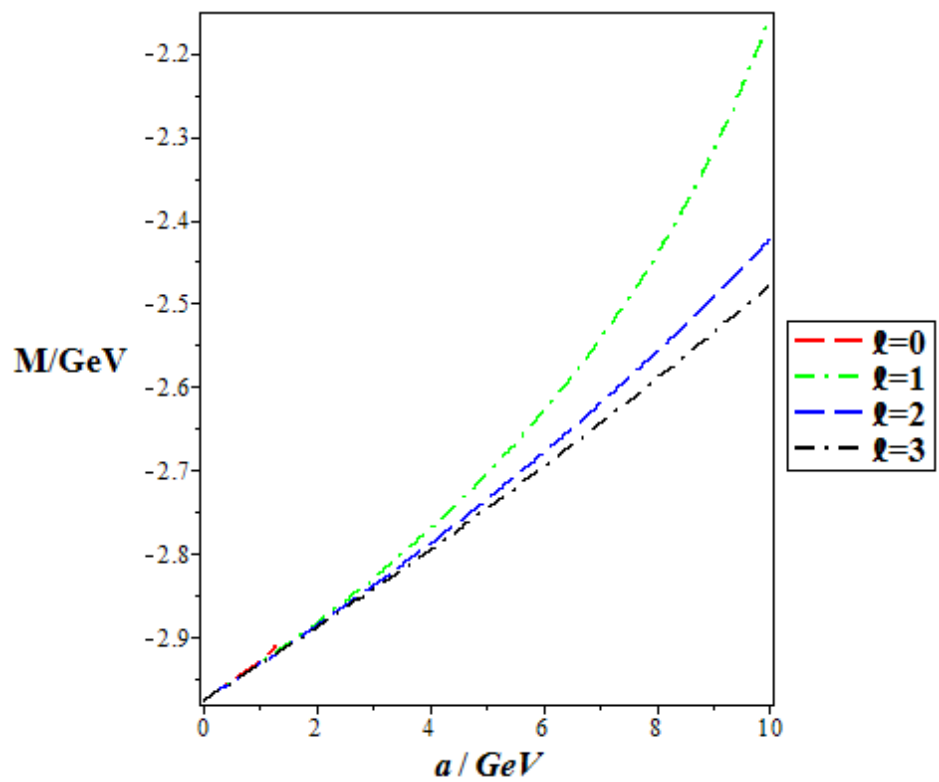

Fig. 2. Variation of mass spectra with potential strength $(a)$ for different quantum numbers

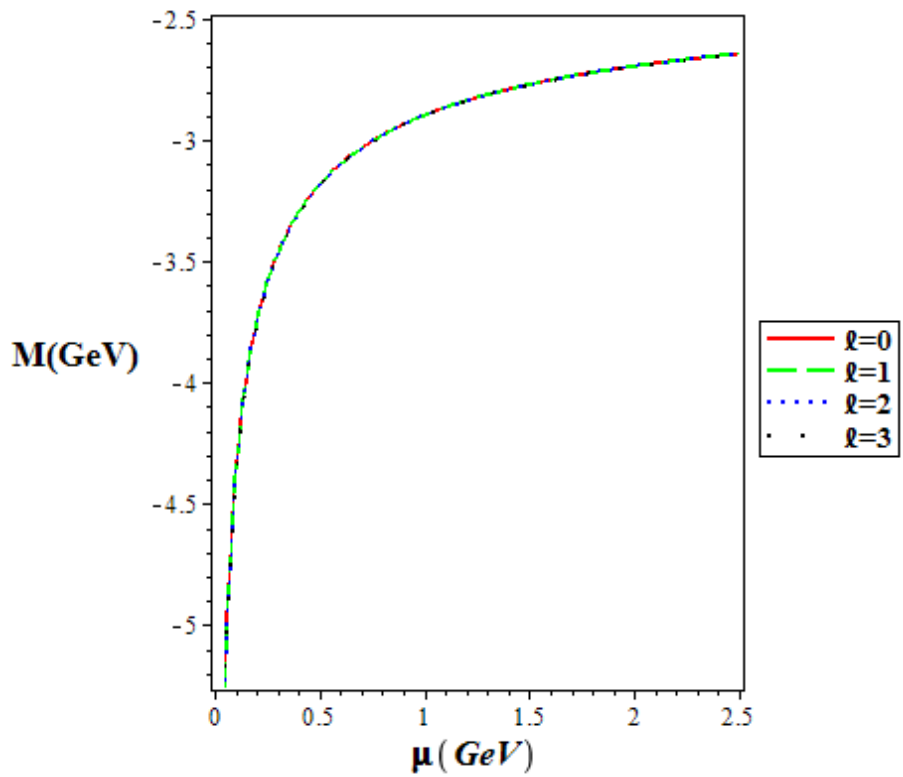

Fig. 3. Variation of mass spectra with reduced mass $\mu$ for different quantum numbers

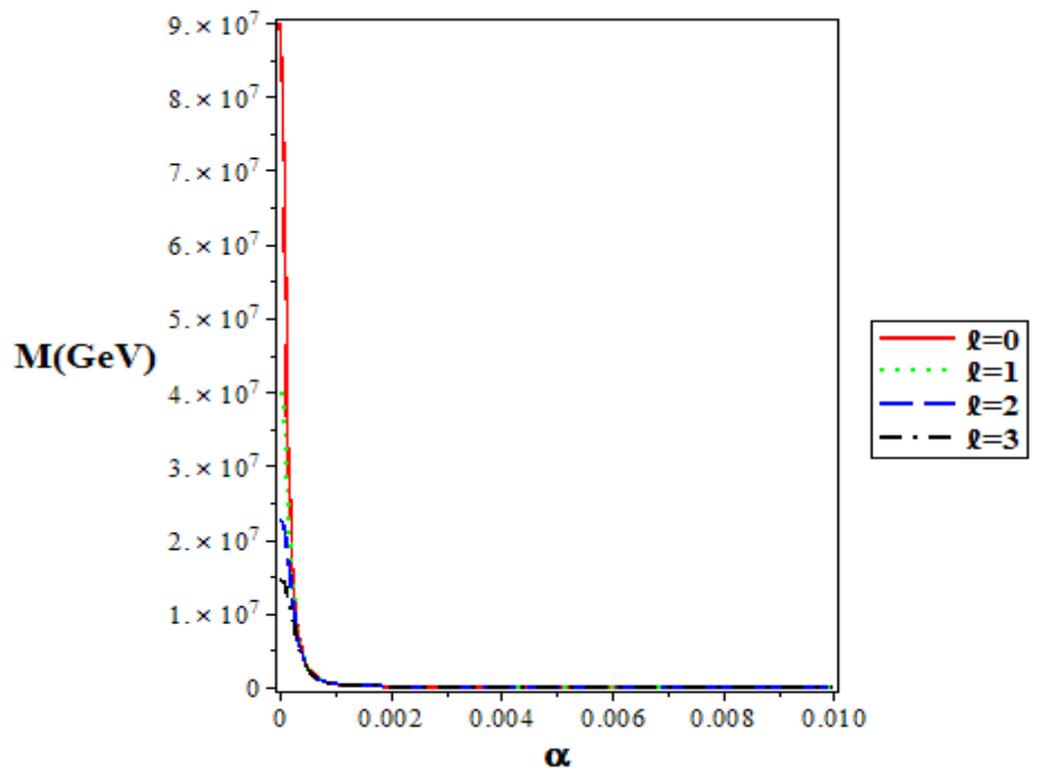

Fig. 4. Variation of mass spectra with screening parameter ($\alpha$) for different quantum number

## 4.2 Discussion of results

We calculate mass spectra of charmonium and bottomonium for states from 1S to 1F by using Eq. (46). We adopt the numerical values of bottomonium $(b\bar{b})$ and charmonium $(c\bar{c})$ masses as $4.68\ GeV$ and $1.488\ GeV$, respectively, Ref. [34]. Then, the corresponding reduced mass are $\mu_b = 2.340\ GeV$ and $\mu_c = 0.744\ GeV$. The free parameters of Eq.(46) were then gotten by solving two algebraic equations by inserting experimental data of mass spectra for $2S, 2P$ in the case of charmonium. In the case of bottomonium the values of the free parameters in Eq. (46) are calculated by solving two algebraic equations, which were obtained by inserting experimental data of mass spectra for $1S, 2S$. Experimental data is taken from Ref. [35].

We note that calculation of mass spectra of charmonium and bottomonium are in good agreement with the experimental data given that the maximum error in comparison with the experimental data is 0.0055 $GeV$. The values obtained are in good agreement with the works of other researchers like in Ref.[30] and Ref.[20], as shown in tables 1 and 2. In Ref.[30] the author investigated the N-radial SE analytically by employing Cornell potential, which was extended to finite temperature. In Ref.[20] the Klein-Gordon equation is solved for the Yukawa potential using the Nikiforov-Uvarov method. The energy eigenvalues were obtained both in relativistic and non-relativistic regime. The results were used to calculate heavy-meson masses of charmonium $c\bar{c}$ and bottomonium $b\bar{b}$. We also plotted mass spectra energy against potential strength $(a)$, reduced mass $(\mu)$ and screening parameter $(\alpha)$ respectively. In Fig. 2, the mass spectra energy converges at the beginning but spread out and there is a monotonic increase in potential strength $(a)$. Figures 3 and 4 shows the convergence of the mass spectra energy as the screening parameter $(\alpha)$ and reduced mass $(\mu)$ increases for various angular momentum quantum numbers. This indicates the energy peak as an observable to determine the top quark mass.

## 5. Conclusion

In this work, the bound state solutions of the Schrodinger equation for the Varshni potential using the Nikiforov-Uvarov method were obtained. The corresponding eigenfunction was achieved in terms of Laguerre polynomials. We applied the present results to calculate heavy-meson masses such as charmonium and bottomonium. The energy eigenvalues of charmonium $(c\bar{c})$, and bottomonium $(b\bar{b})$ for states 1S to 1F were obtained and compared with experimental data and other theoretical works, which are in good agreement with the maximum error of $0.0055\ GeV$. The exponential type potential has been successfully applied in predicting the mass spectra of heavy mesons. The analytical solutions can also be used to describe other characteristics of the quarkonium systems like thermodynamic properties.